# Colossal electroresistance in metal/ferroelectric/semiconductor tunnel diodes for resistive switching memories

Zheng Wen, Chen Li, Di Wu*, Aidong Li and Naiben Ming

*National Laboratory of Solid State Microstructures and Department of Materials Science and Engineering, College of Engineering and Applied Sciences, Nanjing University, Nanjing 210093, China*

*Corresponding author: diwu@nju.edu.cn

**Abstract**

We propose a tunneling heterostructure by replacing one of the metal electrodes in a metal/ferroelectric/metal ferroelectric tunnel junction with a heavily doped semiconductor. In this metal/ferroelectric/semiconductor tunnel diode, both the height and the width of the tunneling barrier can be electrically modulated due to the ferroelectric field effect, leading to a colossal tunneling electroresistance. This idea is implemented in Pt/BaTiO$_3$/Nb:SrTiO$_3$ heterostructures, in which an ON/OFF conductance ratio above $10^4$ can be readily achieved at room temperature. The colossal tunneling electroresistance, reliable switching reproducibility and long data retention observed in these ferroelectric tunnel diodes suggest their great potential in non-destructive readout nonvolatile memories.

Ferroelectric tunnel junctions (FTJs), composed of two metal electrodes separated by an ultrathin ferroelectric barrier, have attracted much attention as promising candidates for nonvolatile resistive memories. Theoretic **[1-4]** and experimental **[5-9]** works have revealed that the tunneling resistance switching in FTJs originates mainly from a ferroelectric modulation on the tunneling barrier height. In quantum mechanics, the wave function of an electron can leak through a barrier as long as the barrier height is sufficiently low and the barrier width is sufficiently thin. The electron therefore has a finite probability of being found on the opposite side of the barrier.**[10]** Conventionally, a tunnel junction consists of two metal electrodes and a nanometer-thick insulating barrier layer sandwiched inside. The transmittance of a tunnel junction depends exponentially on the height and the width of the barrier.**[10]** Superconductor Josphson junctions **[18,19]** and magnetic tunnel junctions **[20,21]** are famous examples that have attracted much attention. Recently, great technological advances have been achieved in growth of perovskite oxide thin films. Ferroelectricity is found to exist even in thin films of several unit cells (u.c.) in thickness, which makes it possible to realize ferroelectric tunnel junctions (FTJs) by employing ultrathin ferroelectrics as the barriers.**[22,23]**

The idea of FTJs can be dated back to the early 1970s, when Esaki *et al.* reported a possible FTJ using bismuth niobate as the barrier.**[24]** However, the control mechanism of FTJs was formulated only very recently.**[2-4]** Unlike the conventional tunnel junctions, the barrier height seen by electrons in a FTJ can be modulated electrically by polarization reversal in the ferroelectric barrier.**[2-4]** This results in an electrical switching of tunneling resistance, the so-called tunneling electroresistance (TER). The electrically modulated barrier height have been demonstrated by recent experiments.[5-9] Gruverman *et al.* verified that the resistance switching in ultrathin $BaTiO_3$(BTO)/$SrRuO_3$/$SrTiO_3$ epitaxial heterostructures is associated with barrier height modulation in response to polarization reversal by scanning probe microscopy.[7] Giant TER with reliable switching has been reported by Garcia *et al.* [6] and Pantel *et al.* [9] in FTJs using BTO and $Pb(Zr,Ti)O_3$ as barriers. These achievements suggest that the FTJ offers a promising alternative for nonvolatile resistive memories of pure electronic contribution.[6] Moreover, nonvolatile switching between four tunneling resistance states has been realized in multiferroic tunnel junctions.[13-17] Most recently, a ferroelectric memristor effect has also been demonstrated by carefully controlling the domain switching process in FTJs.[25] However, in addition to the barrier height, the tunneling transmittance also depends exponentially on barrier width according to basic quantum-mechanics.[10] Dramatically enhanced TER can be expected if the barrier width is also electrically modulated in parallel with the barrier height.

In this letter, we propose a ferroelectric tunnel diode employing a metal/ferroelectric/semiconductor heterostructure, as shown in Fig. 1, using an n-type semiconductor for example. By polarization reversal in the ultrathin ferroelectric barrier, the semiconductor surface can be switched between accumulation and depletion of majority carriers due to a ferroelectric field effect.[11,12] Hence, along with the switching of barrier height in response to the polarization reversal in the ferroelectric barrier, there exists an additional tuning on the width of the barrier since the tunneling electrons have to experience an extra barrier over the space charge region if the semiconductor surface is depleted.

As depicted schematically in Fig. 1a, if the ferroelectric polarization points to the semiconductor, positive bound charges in the ferroelectric/semiconductor interface shall drive the n-type semiconductor surface into accumulation.[11,12] The accumulated semiconductor can be treated as a metal and the screening is then similar to that in metal/ferroelectric/metal FTJs.[2-4] The screening is usually incomplete and a depolarization field opposite to the polarization develops in the ferroelectric barrier.[3] This depolarization field lowers the barrier height and generates a higher tunneling transmittance.[2-4] The device is then set to a low resistance ON state. However, as the polarization is reversed, pointing to the metal electrode as shown in Fig. 1b, the semiconductor surface is depleted of electrons and the negative ferroelectric bound charges have to be screened by the ionized donors.[11,12] Contrast to the ON state where majority carriers can be accumulated very close to the ferroelectric/semiconductor interface, the immobile screening charges in the depleted state spread over a space charge region defined by the doping profile.[26] On one hand, the incomplete screening again produces a depolarization field, but this time increases the barrier height.[2-4] Note that the screening length of the ionized donors in the depleted state can be orders larger than that of the electrons in the accumulated state, the depolarization field and the modulation on the barrier height can be much stronger in the depleted state.[3] On the other hand, the tunneling electrons have to experience an extra barrier in the depleted space charge region, due to band bending induced by ferroelectric polarization in the barrier.[26] Estimated using reasonable parameters, the width of the space charge region can at least be several nanometers, comparable to the width of the barrier. (Supplementary) Therefore, the tunneling transmittance can be dramatically decreased by this extra barrier. The device is then set to a high resistance OFF state. Compared with the metal/ferroelectric/metal FTJs, the metal/ferroelectric/semiconductor diodes may have a greatly enhanced TER due to the extra barrier in the semiconductor. In other words, the proposed ferroelectric tunnel diodes can also be regarded as FTJs in which the screening length of one of the electrodes can be tuned by the polarization in the barrier. The greatly enhanced TER is a result of the enhanced contrast in screening length.[3]

We calculated the conductance in the ON and the OFF states for a model ferroelectric tunnel diode (Supplementary), following the method implemented by Zhuravlev and coworkers.[3] The TER value, as quantified by the calculated ON/OFF conductance ratio, depends strongly on the OFF state resistance, which increases abruptly when the polarization exceeds a threshold corresponding to the critical polarization that could bend the conduction band minimum above the Fermi level to form the extra barrier. The TER value can reach $10^4 \sim 10^5$ for a moderate polarization of 10~15 μC/cm$^2$, at least 2 orders larger than that in FTJs.[3] (Supplementary Fig. S2)

Single-crystalline 0.7wt% Nb-doped SrTiO$_3$ (Nb:STO) is chosen as the semiconductor substrate to facilitate epitaxial growth of the ultrathin BTO ferroelectric barrier.[27] A 7 u.c. thick BTO ultrathin film was deposited epitaxially on (001) oriented Nb:STO substrates by pulsed laser deposition in a typical layer-by-layer growth mode. (Supplementary Fig. S3a) The BTO/Nb:STO heterostructures exhibit atomically flat surfaces with the root-mean-square roughness about 0.1 nm over an area of 5×5 μm$^2$. (Supplementary Fig. S3b) As shown in Fig. 2a, reciprocal space mapping around the (013) Bragg reflection indicates coherent cube-on-cube growth of BTO on Nb:STO substrates. The measured out-of-plane and in-plane lattice constants are 0.42 and 0.39 nm, respectively. The tetragonality (c/a ratio) of 1.08 implies a high Curie temperature and a large ferroelectric polarization in the ultrathin BTO.[23,28] Ferroelectricity of BTO is characterized by pizeoresponse force microscopy (PFM). A triangle wave of 5.0 V in amplitude is used to detect the local PFM response. The hysteresis loops are shown in Fig. 2b, which indicates the ferroelectric nature of the 7 u.c. BTO on Nb:STO. The coercive voltages are about +2.0 and -1.5 V indicated by the minima of the amplitude loop. Fig. 2c shows an out-of-plane PFM phase image of ferroelectric domains written on the BTO surface. It is observed that the as-grown BTO thin film exhibits a preferred polarization direction pointing towards the film surface. By scanning a DC-biased Pt-coated tip on the BTO/Nb:STO surface, the as-grown BTO can be written into fully polarized domains. The 180$^0$ phase contrast reveals that the polarization is antiparallel in the two domains.

Pt top electrodes of 30 μm in diameter were deposited on BTO/Nb:STO heterostructures to form ferroelectric tunnel diodes. (Supplementary Fig. S4) Write pulses of various amplitudes in both polarities, as illustrated in Fig. 3a, are used to switch ferroelectric polarization in the BTO barrier. Corresponding tunneling resistance is read by pulses of +0.1 V in amplitude following each write pulse. We will show that the resistance switching characteristic of the Pt/BTO/Nb:STO devices is in good agreement with the proposed mechanism. The room temperature tunneling resistance as a function of write amplitude is shown in Fig. 3b. The as-deposited Pt/BTO/Nb:STO tunnel diode exhibits a high tunneling resistance, consistent with the preferential polarization direction in the as-grown BTO thin film, as shown in Fig. 2c. With small amplitudes, the polarity reversal of the write pulses cannot modify the resistance states in Pt/BTO/Nb:STO since the amplitude is too small to switch the ferroelectric polarization in the BTO barrier. When the amplitude of write pulses are comparable to the coercive voltages, the polarization in BTO barrier starts to reorient in response to the external electric field. Hence, the low resistance ON state (accumulation of Nb:STO surface) is established abruptly and does not change much with the increases of write amplitude, in consistent with the result of calculation. (Supplementary Fig. S2) In contrast, the OFF state resistance increases with increasing write amplitude above 1.8 V in response to the development of the depleted space charge region in the semiconductor. As a consequence, the TER increases rapidly with the increase of write amplitude. The ON/OFF ratio at a write amplitude of 3.5 V can reach about 13,000. Corresponding current-voltage (I-V) characteristics of the ON and the OFF states are shown in Fig. 3c. The I-V curves are highly non-linear in both ON and OFF states and the current contrast of about 4 orders can be easily observed. The current densities are $2.55\times10^4$ and 2.12 mA/cm$^2$ for the ON and the OFF states, respectively, at 0.1 V. The best reported room temperature ON/OFF ratio (~100) of FTJs using ultrathin BTO as the barriers, appears in Au/Co/BTO/(La,Sr)MnO$_3$ heterostructures.[6] In comparison with the I-V data in Au/Co/BTO/(La,Sr)MnO$_3$,[6] the OFF state current density of Pt/BTO/Nb:STO is about 2 orders smaller, while the current density of the ON state is comparable. This suggests that the proposed ferroelectric tunnel diode may be more effective to shut off the current in the OFF state, due to the formation of the extra barrier in the space charge region.

Chang *et al.* proposed a metal/ferroelectric/semiconductor tunneling structure in the 1970s.[29] Assuming there exist high density interfacial states at the ferroelectric/semiconductor interface, the TER comes from a ferroelectric modulation of Schottky barrier heights due to charge/discharge of the interfacial states. The mechanism proposed in the present work is significantly different from the previous one. Unlike ferroelectric/semiconductor combinations such as evaporated SbSi on Si and chemically formed KH$_2$AsO$_4$ on GaAs, proposed by Chang and Esaki,[29] the high quality epitaxial deposition of ultrathin ferroelectrics on lattice matched oxide semiconductors greatly suppresses the interfacial states.[11] The TER then originates mainly from the ferroelectric modulation on the space charge region in the semiconductors[11,12], which the electrons have to tunnel through. As shown in Fig. 1, the Pt/BTO/Nb:STO tunneling device can be regarded as the capacitor of the BTO barrier connected in series with the capacitor of the space charge region in Nb:STO semiconductor. When the semiconductor surface is accumulated, the capacitance of the space charge region is negligible in the ON state. The measured capacitance of the OFF state should be smaller than that of the ON state due to the existence of the space charge region capacitance. This reduction of capacitance in the OFF state is indeed observed, indicating the existence of a space charge region in Nb:STO. (Supplementary Fig. S5)

Data retention and switching properties are the most important reliability issues associated with ferroelectric memory devices. Fig. 3d shows the retention property of the Pt/BTO/Nb:STO tunnel diode. There is no significant reduction in the resistance contrast before and after a 24-hour retention time as shown in the inset. An ON/OFF ratio over 3 orders can still be retained by extrapolating the retention data to over 10 years. These suggest the excellent resistance stability of the ferroelectric tunnel diodes during repetitive readouts and long time retention. The switching characteristic of the Pt/BTO/Nb:STO diodes was tested by switching the ferroelectric barrier repetitively with ±2.5 V pulses. As shown in Fig. 3e, a typical bipolar resistance switching with good reproducibility is achieved. The ON/OFF ratio of ~100 is maintained over 3,000 write/read cycles at room temperature.

In summary, we have proposed a metal/ferroelectric/semiconductor heterostructure that may greatly enhance the nonvolatile memory effect employing tunneling through a ferroelectric barrier. The colossal electroresistance observed is ascribed to the creation/elimination of an extra barrier on the semiconductor surface in response to the polarization reversal in the ferroelectric barrier. The characteristics such as colossal electroresistance, long retention and good switching reproducibility make the proposed scheme a promising candidate for nonvolatile resistive memories. Moreover, if a ferromagnetic semiconductor and/or a multiferroic barrier are used, the idea presented here can be readily extended into multiferroic tunneling devices, where four or more nonvolatile states with great resistance contrast may be realized in a single memory node.**[13-17]**

**Acknowledgements**

This work was jointly sponsored by State Key Program for Basic Research of China (2009CB929503) and Natural Science Foundation of China (91022001 and 51222206). Shanghai Synchrotron Radiation Facility is greatly acknowledged for providing the beam time and technical assistance.

**Figure captions**

**Figure 1. Resistive switching mechanism in a ferroelectric tunnel diode.** Schematic drawings of the metal/ferroelectric/semiconductor structures and corresponding potential energy profiles for the low (**a**, **c**) and the high (**b**, **d**) resistance states. In **a** and **b**, M, F, and S stands for the metal, the ferroelectric and the semiconductor, respectively. A heavily doped n-type semiconductor is taken for example. In the ferroelectric barriers, the red arrows denote the polarization directions and the **+**/**−** symbols represent positive/negative ferroelectric bound charges. The **+**/**•** symbols in the metal and n-type semiconductor electrodes represent holes/electrons. The ⊕ symbols represent ionized donors. A rectangular barrier, denoted by dashed lines in **c** and **d**, is assumed when the ferroelectric is un-polarized. To simplify, we further assume that ferroelectric bound charges at the metal/ferroelectric interface can be perfectly screened. The barrier height at metal/ferroelectric interface is, therefore, fixed and does not change with the polarization reversal.

**Figure 2. Structure and ferroelectricity. a**, Reciprocal space mapping around the (013) Bragg reflection of BTO/Nb:STO heterostructures. **b**, Local PFM hysteresis loops: top, phase signal; bottom, amplitude signal. **c**, PFM out-of-plane phase image after writing in an area of 3×3 $\mu m^2$ by +5 V and then the central 1.5×1.5 $\mu m^2$ by -5V.

**Figure 3. TER of Pt/BTO/Nb:STO ferroelectric tunnel diodes at room temperature. a**, Schematic drawing of a period of the test pulse train and the corresponding polarization directions in the BTO ferroelectric barrier for the two resistance states (ON, low resistance state; OFF, high resistance state). **b**, Resistance of the ON and the OFF states and the ON/OFF conductance ratio as a function of write amplitude. **c**, Current-voltage curves at the ON and OFF states. **d**, Retention properties. The inset shows current-voltage curves collected before (as-written) and after 24 hours

retention measurement. **e**, Bipolar resistance switching between the ON and OFF states.

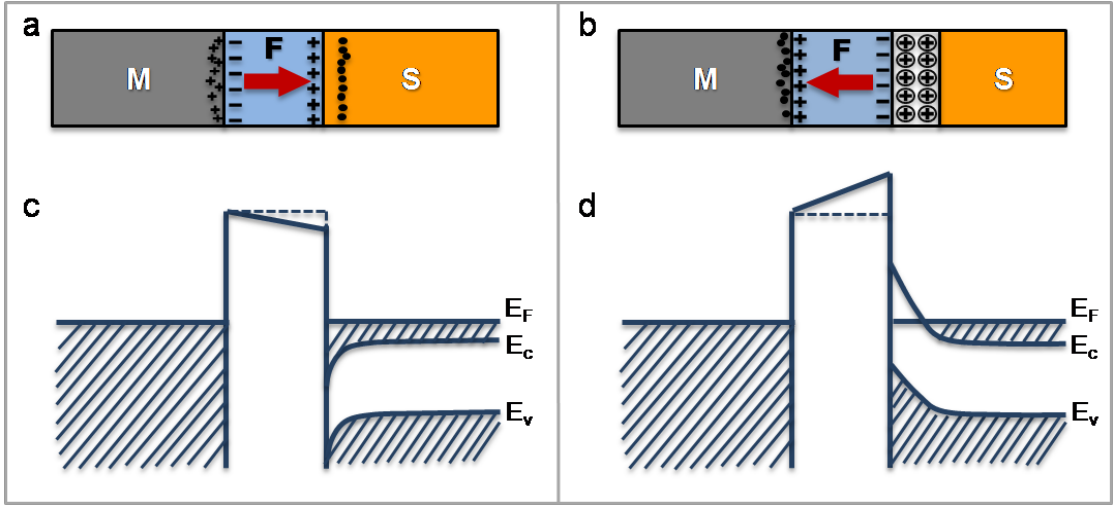

**Figure 1**

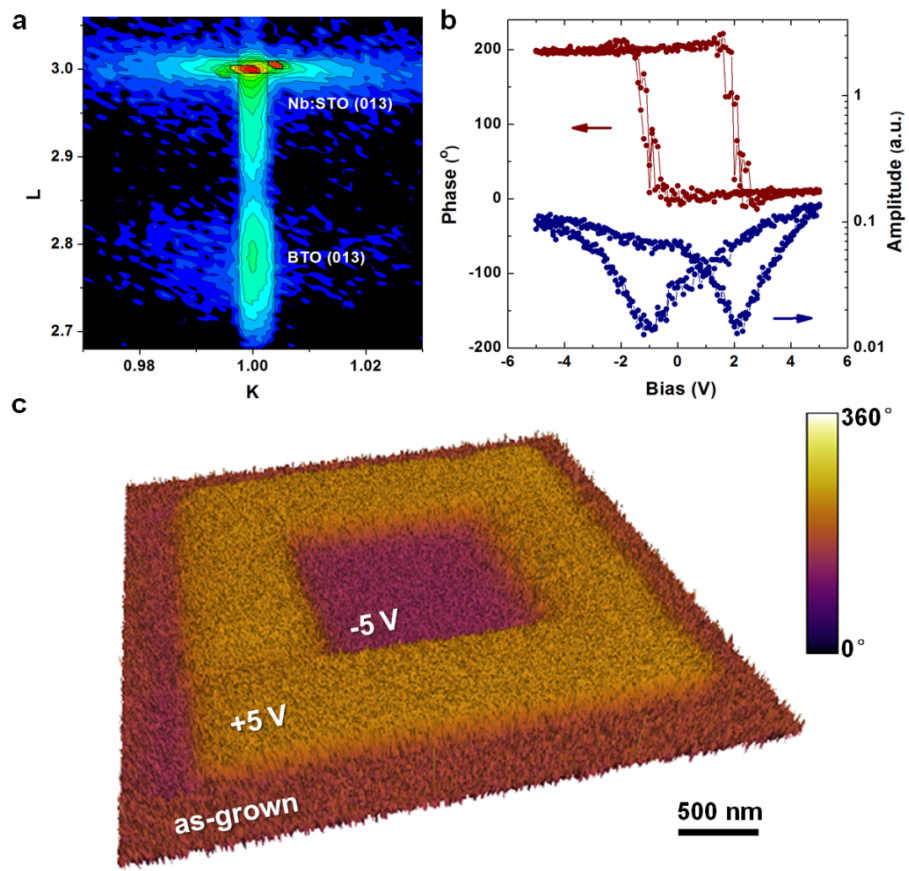

**Figure 2**

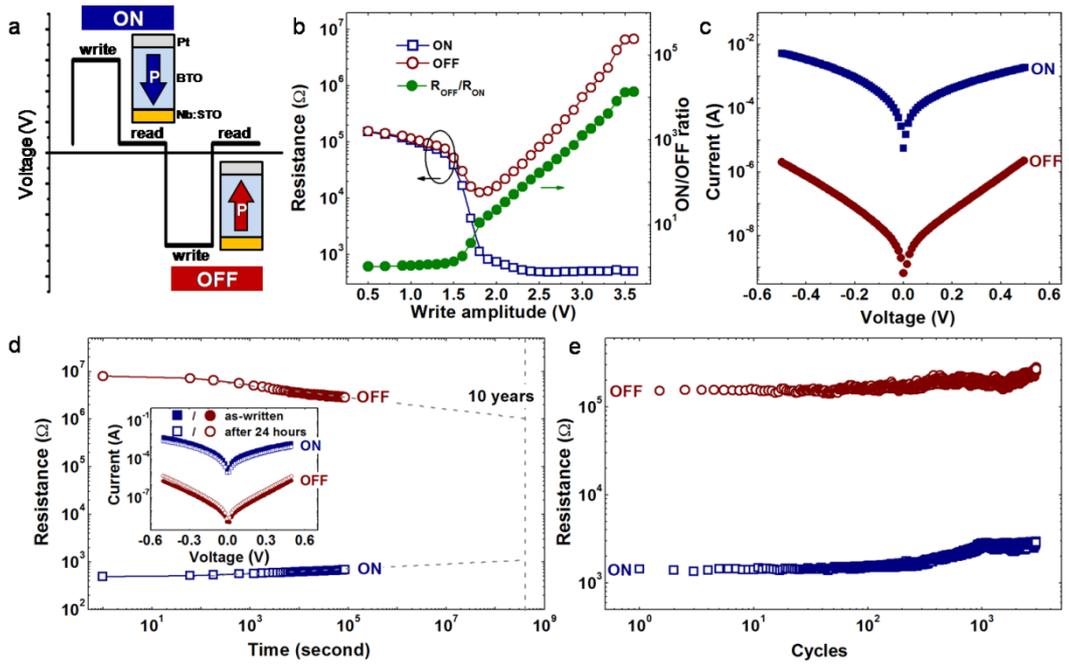

**Figure 3**